**Evidence for a charged vortex lattice in an exciton superfluid**

Ryan L. Lee[1]*, Jungwoo Lee[1]*, Viliam Vaňo[1]*, Bo Zou[2], Cheng-Li Chiu[1], Kenji Watanabe[3], Takashi Taniguchi[4], Allan MacDonald[2], and Ali Yazdani[1†]

[1] *Joseph Henry Laboratories and Department of Physics, Princeton University, Princeton, NJ, 08544, USA*

[2] *Department of Physics, University of Texas at Austin, Austin, TX, 78712, USA*

[3] *Research Center for Functional Materials, National Institute for Materials Science, 1-1 Namiki, Tsukuba 305-0044, Japan*

[4] *International Center for Materials Nanoarchitectonics, National Institute for Materials Science, 1-1 Namiki, Tsukuba 305-0044, Japan*

* These authors contributed equally to this work.

† Corresponding author email: yazdani@princeton.edu

**Lattices of quantized vortices are a defining property of rotating superfluids and of type-II superconductors in magnetic fields[1–3]. Realizing an analogous vortex lattice in an exciton superfluid—formed from bound electron–hole pairs—has remained challenging because excitons are electrically neutral. Building on previous theoretical studies of the low-energy excitations in an exciton superfluid[4–6], recent theory has predicted that in a strong magnetic field, a weakly charged exciton superfluid can stabilize a vortex–antivortex lattice arising from the mutual repulsion of fractionally charged vortex cores[7]. Here we report evidence for this predicted vortex lattice in a bilayer exciton condensate by imaging its charged vortices with the scanning tunnelling microscope. The condensate forms between Landau levels of two graphene layers separated by a thin hexagonal boron nitride barrier in a perpendicular magnetic field[8–10]. We find that the vortex lattice periodicity scales with the total excess charge and is insensitive to its distribution between the two layers, in quantitative agreement with theoretical predictions. Our observations provide evidence for phase-coherence in an exciton superfluid system.**

Fermionic systems with attractive interactions can undergo pairing to form composite bosons, such as Cooper pairs of electrons or excitons composed of bound electron–hole pairs[11]. Optically generated excitons, however, typically have short lifetimes owing to rapid electron–hole recombination[12–14], and the electrical neutrality of direct excitons makes them difficult to probe through transport measurements. By contrast, equilibrium interlayer excitons, with electrons and holes in separate layers at matched densities, circumvent these limitations[8,9,15–19]. Their spatial separation greatly prolongs exciton lifetimes, while independent electrical contacts to each layer enable direct transport probes and gate-tuneable control of exciton density. Their interlayer character also produces repulsive exciton–exciton interactions and strong correlations[20,21]. Interlayer excitons have been realized in the Landau levels of double-layer GaAs systems[15,16,22–27], graphene[8–10,28–30], and transition-metal dichalcogenides[12,17–19,21,31–33].

At sufficiently low temperatures, interlayer excitons condense into a Bose–Einstein condensate, developing interlayer phase coherence that spontaneously breaks a U(1) symmetry through Coulomb interactions, without requiring interlayer tunnelling[11,34,35]. This exciton condensate is a macroscopically coherent quantum state, analogous to a BCS superconductor but with a simpler and highly tuneable pairing mechanism that has been proposed as a platform for dissipation-less transport[9]. Experimental signatures include perfect Coulomb drag[8,16,17,19,23,36,37], vanishing counterflow resistance[9,26,29,38], a zero-bias interlayer tunnelling anomaly[22,25,30,39], and spatial photoluminescence coherence measurements[31], although unambiguous evidence of global phase coherence has remained elusive.

The condensate can be described as a layer-pseudospin ferromagnet, whose out-of-plane component represents layer polarization and whose in-plane component encodes the relative interlayer phase[40]. The lowest energy charged excitations of this condensate are vortices and antivortices in the U(1) exciton superfluid order parameter[6,7,41,42]. In contrast with vortices in type-II superconductors, the vortices in an exciton condensate carry charge instead of magnetic flux. Phase stiffness energetically favours zero net vorticity, and additional charge carriers are predicted to fractionalize and localize at the cores of equal numbers of vortices and antivortices (see SI section 1). Although vortices and antivortices typically annihilate, long-range Coulomb repulsion between their fractionally charged cores can stabilize a honeycomb lattice in which vortices and antivortices with opposite layer polarization occupy opposite sublattices[7]. Although theoretical studies of vortex crystal phases in bilayer quantum Hall systems are well-established[5,43], the honeycomb symmetry was predicted only recently[7]. In this state, each sublattice carries fractional charge of the same sign, yielding one elementary charge per unit cell[7]. Fig. 1a illustrates such a lattice, showing simulated charge densities in the top and bottom graphene layers and the layer pseudospin. Charge maxima in the top (bottom) layer correspond to vortices (antivortices), shown in red (blue). At the core of each vortex/antivortex, the exciton amplitude vanishes and the wavefunction is fully polarized in a single layer. At the centre of each hexagon, the condensate amplitude is peaked, with its phase indicated by black arrows.

Here we search for this predicted vortex phase in a quantum Hall double-layer graphene system by directly imaging the charged cores of its vortices using scanning tunnelling microscopy (STM). Because excess charge localizes at vortex cores, spatial modulations of the tunnelling density of states image the vortex lattice. Quantitative comparison with theoretical predictions for the lattice structure, and clear distinction from other charge-ordered phases such as a Wigner crystal, are enabled by systematic measurements as a function of carrier density, carrier type, and interlayer density imbalance. Moreover, the lattice spacing depends only on the total filling factor across both layers—a hallmark of interlayer coherence. The observation of a vortex lattice provides compelling evidence for the phase-coherent nature of the exciton condensate.

## Spectroscopy of an excitonic condensate and imaging of a vortex lattice

The devices used for our experiments consist of two monolayer graphene layers separated by thin (~3.2 nm) hexagonal boron nitride (hBN) and placed on a graphite backgate, with the STM tip probing the top graphene layer. At 9 T, the ratio of interlayer spacing to magnetic length is ~0.37, putting the system in the Bose-Einstein condensate (BEC) regime[9]. As shown in Fig. 1b, the device geometry allows independent tuning of the total filling factor $\nu_{total} = \nu_{top} + \nu_{bot}$ with the backgate voltage $V_g$ and redistribution of carriers between the layers with the interlayer voltage $V_{int}$. In a perpendicular magnetic field, graphene's Dirac spectrum forms fourfold spin- and valley-degenerate Landau levels (LL). Electron-electron interaction lifts this degeneracy, leading to quantum Hall ferromagnetic (QHFM) states in each layer[40,44]. When the layers are sufficiently close and both LLs are partially filled such that $\nu_{total}$ is an integer, spontaneous interlayer coherence emerges, giving rise to an interlayer integer quantum Hall effect[24]. Equivalently, under a particle-hole transformation in one of the layers, this state can be viewed as a bilayer exciton condensate with equal electron and hole filling factors.

Figure 1c shows the tunnelling gap of the top graphene layer as a function of $V_g$ and $V_{int}$, measured at magnetic field $B = 9$ T and temperature $T = 200$ mK. Variations in this gap distinguish compressible and incompressible regimes of the top layer and closely match the simulated electrostatic phase diagram in Fig. 1d (see SI section 2)[45]. We attribute the enlarged gaps at the top and bottom of the plotted range to the $\nu_{top} = 0$ and $\nu_{top} = -1$ QHFM of the top layer (green region in Fig. 1d). At $V_g = -1.9$ V and 10 mV < $V_{int}$ < 30 mV, the system enters a regime with $\nu_{total} = -2$, while the individual layers remain at non-integer filling. In this regime, the electron density in one layer precisely matches the hole density in the other, and a pronounced spectroscopic gap signals the formation of an interlayer exciton insulator[45]. The magnitude of this gap varies systematically with $V_{int}$, exhibiting a minimum of approximately 40 mV near half filling of both layers' Landau levels ($V_{int} \sim 23$ mV) where the exciton density is expected to be largest (see SI section 3). This behaviour is consistent with density-dependent screening of the excitonic interaction. The fourfold LL degeneracy permits 16 combinations of zeroth LL flavours between the layers, yielding 16 possible exciton condensates (Fig. S2). In this work, we focus on the exciton condensate with $\nu_{total} = -2$, $-1 < \nu_{top} < 0$ and $-2 < \nu_{bot} < -1$.

Evidence for a charged vortex lattice emerges upon doping the exciton condensate. Spatially resolved measurements of the tunnelling current $I(V_b, \boldsymbol{r})$ (Figs. 1e,f) reveal emergent periodic structures for both electron and hole doping at the conditions marked in Fig. 1d, where both layers are compressible. In this regime, transport measurements on double layer graphene have revealed perfect Coulomb drag and zero counterflow resistance[8,9]. Our maps are acquired over extended defect-free regions, with $V_b$ set near the gap edge and using work-function-matched tips (see Methods and SI sections 4 and 5). As in type-II superconductors, a vortex lattice extending over the ~400 nm field of view provides a lower bound of similar size on the length scale of phase coherence in the exciton condensate (see SI section 6). The resulting real-space images and the extracted modulus of the structure factor $|S(q)|$ display clear translational symmetry breaking with sixfold Bragg peaks, demonstrating the emergence of a charge lattice. The lattices are slightly distorted by local disorder. As shown below, their properties quantitatively agree with predictions for a charged vortex lattice and clearly distinguish it from alternative charge-ordered phases, such as a Wigner crystal of electrons or holes.

## Evolution of the vortex lattice with doping

A key characteristic of the charged vortex lattice is its evolution with net doped charge density[7]. This dependence tests whether the periodicity is set by the density of charged vortices and probes the predicted transition from a vortex lattice to a conventional Wigner crystal at higher doping[7]. Figure 2a shows that the spectroscopic exciton gap at the Fermi level ($E_F$) (near $V_g \approx -1.9$ V) evolves into a residual Coulomb gap upon electron doping (see SI section 7). The Coulomb gap corresponds to the interaction-driven suppression of local tunnelling density of states around $E_F$, widely observed in high-field tunnelling spectroscopy of partially filled Landau levels[46–52]. Similar transitions between incompressible and compressible states have been reported in tunnelling studies of partially filled Landau levels[49,50,52]. At finite doping, the spectroscopic gap develops clear spatial modulations (Fig. 2b), which provide the contrast in the real-space $I(\boldsymbol{r})$ maps of the electron-doped condensate (Fig. 2c–h). The real space maps are taken with a tunnelling setpoint bias of $V_b = -100$ mV at a current of I = 400 pA. These spatial modulations stem from two mechanisms. The first is a varying residual exciton gap above vortex cores, analogous to that observed in STM studies of type-II superconductors[53]. The second is the Coulomb interaction between charged vortices and tunnelling electrons, which shifts the Coulomb gap as in prior STM studies of Wigner crystals[54] (see SI section 8.1). These measurements reveal a lattice of bright features that becomes progressively denser with increasing carrier density and is locally pinned and distorted by subsurface defects (see SI section 9).

We compute the modulus of the structure factor $|S(q)|$ from the real-space $I(\boldsymbol{r})$ images to extract the lattice spacing (see Methods and SI section 10)[54]. The images reveal a continuous evolution: from a sparse and partially disordered lattice at low doping, to a well-ordered lattice with six distinct Bragg peaks at intermediate doping, and ultimately to a disordered or melted phase exhibiting ring-like features in $|S(q)|$ at higher doping (see SI section 20). Similar behaviour has been reported for disordered vortex lattices in type-II superconductors[55]. The extracted lattice constant (Fig. 2o, black symbols) closely agrees with the predicted vortex lattice constant $a = \sqrt{4\pi l_B^2/\sqrt{3}\tilde{\nu}_{total}}$ (red curve), where $l_B$ is the magnetic length and $\tilde{\nu}_{total}$ is filling relative to the neutral exciton condensate. The spatial pattern observed in *I(**r**)* within a unit cell is triangular lattice-like, not honeycomb-like, suggesting that tunnelling from occupied states peaked on the honeycomb sublattice whose charge is localized in the top layer. This is because in the probed range of $V_{int}$, top-layer vortices carry more charge than those in the bottom layer and dominate the signal.

An alternative explanation is an electron Wigner crystal formed solely in the top graphene layer, whose lattice constant would be determined exclusively by the top-layer filling factor. Near the exciton gap, however, the bottom layer remains compressible and screens the electric field from the backgate, keeping the top-layer filling factor nearly constant. This is inconsistent with our observation of a continuously varying lattice constant as $V_g$ approaches the exciton gap. The lattice constant predicted for a top-layer Wigner crystal (Fig. 2o, blue dashed line) clearly disagrees with the data for $\tilde{\nu}_{total} < 0.17$. Defect scattering is ruled out in SI section 11.

The vortex lattice is expected to transition into a conventional Wigner crystal once one of the layers approaches integer filling[7]. In our measurements, this transition occurs near $\tilde{\nu}_{total} \sim 0.17$ ($V_g \sim -1.74$ V), where both the experimental spectroscopy (Fig. 1c) and the calculated phase diagram (Fig. 1d and inset of Fig. 2o) indicate the onset of an incompressible state. At this point, the measured lattice constant converges to the value expected for a single-layer electron Wigner crystal in the top graphene layer, consistent with the anticipated phase transition[7]. Measurements at a different $V_{int}$ show the same behaviour (see SI section 12).

## Density imbalance evolution

A defining prediction is that the vortex lattice spacing is determined solely by the total excess filling factor across both layers, independent of how charge is distributed between them[7]. This invariance is a direct consequence of interlayer phase coherence: only the total charge couples to the U(1) order parameter. By contrast, the lattice constant of a top-layer Wigner crystal would vary as charge is transferred between layers.

To distinguish between these scenarios, we image the lattice while tuning the interlayer density imbalance using $V_{int}$, keeping the total carrier density fixed (Figs. 3a–f). Over this range, the top layer remains compressible while the bottom layer evolves from incompressible—where a single-layer Wigner crystal would be expected—to compressible, where a vortex lattice of the exciton condensate is anticipated. The lattice symmetry and periodicity remain unchanged as the density imbalance is varied (see SI section 20). Only the image contrast evolves. The structure factors for 12 mV ≤ $V_{int}$ ≤ 22 mV (Figs. 3g–l) exhibit imbalance-independent features and a disorder-induced blurring like that reported for vortex lattices in type-II superconductors[55].

Our electrostatic modelling[45] indicates that, over this range of $V_{int}$, the individual layer filling factors change by up to 30% when both layers are compressible (Fig. 3m). A top-layer Wigner crystal would therefore change its lattice spacing by ~20 nm (Fig. 3n, dashed line). Instead, the experimentally extracted lattice constant (see Methods) remains essentially invariant (Fig. 3n), in quantitative agreement with the vortex lattice prediction (solid line). Microscopically, tuning the density imbalance redistributes fractional charge between vortices and antivortices residing in opposite layers, modifying their layer-resolved charge contrast without altering their spatial positions (see SI section 13). The observed reduction in STM contrast with increasing $V_{int}$ reflects this transfer of charge from top-layer vortices to bottom-layer antivortices (see SI section 8.2 for details).

Together, the invariance of the lattice constant under density imbalance and the systematic evolution of the layer-resolved contrast provide evidence for an interlayer-coherent exciton condensate whose charged excitations are vortices of a U(1) order parameter. When the bottom layer becomes incompressible, the lattice constant converges to that expected for a conventional Wigner crystal in the top layer, consistent with the anticipated phase transition[7]. Furthermore, the observed evolution between compressible and incompressible states rules out alternative explanations such as a Wigner crystal coexisting with an excitonic insulator (see SI section 14). Similar results from different samples, spatial regions, and STM tips are presented in SI section 15.

**Hole-doped vortex lattice**

By particle–hole symmetry, hole doping is expected to produce an analogous vortex lattice[7]. In the electrostatic phase diagram, the hole-doped regime is related to the electron-doped regime by point reflection about ($\nu_{top}$, $\nu_{bot}$) = (-0.5, -1.5), such that excess positive charge should fractionalize into vortex cores in the same manner as excess negative charge. Indeed, upon hole doping we observe a lattice with clear long-range order and six well-defined Bragg peaks (Figs. 4a–l), demonstrating that a vortex lattice forms for both signs of excess charge. Its higher degree of order compared with the electron-doped case likely reflects lower disorder in this region. The tunnelling contrast reverses between the two regimes, consistent with the change in sign of the charge bound to the vortex cores. At negative bias, vortices appear as enhanced tunnelling current under electron doping and as suppressed current under hole doping.

With increasing hole density, the lattice constant decreases (Fig. 4m), mirroring the behaviour on the electron-doped side. Quantitatively, the extracted lattice spacing follows the total excess filling factor across both layers and remains insensitive to interlayer density imbalance (Fig. 4n). This dependence is consistent with the vortex lattice picture and inconsistent with a top-layer hole Wigner crystal. Increasing $V_{int}$, which enhances the hole density in the top layer while keeping the total filling fixed, modifies the STM contrast without changing the lattice spacing, consistent with charge redistribution between vortices and antivortices that reduces the top-layer charge modulation amplitude. Taken together, the symmetric behaviour under electron and hole doping supports the interpretation of the observed charge modulation as a vortex lattice of an interlayer-coherent exciton condensate (see SI section 20).

In analogy with type-II superconductors, the spatial extent of vortex cores provides a measure of the coherence length of the ordered state[56]. By comparing the lattice-site sizes in the vortex lattice and Wigner crystal regimes at fixed lattice constant (see SI section 16), we observe a modest broadening in the vortex phase, consistent with additional charge delocalization around vortex cores. Although the present data do not allow a precise quantitative extraction, the observed trend supports a finite exciton coherence length, which future higher-resolution experiments could measure more precisely.

**Outlook**

Our results establish the formation of a vortex lattice in an interlayer exciton condensate, placing exciton superfluidity on the same conceptual footing as other phase-coherent quantum fluids. The emergence of an ordered array of charged vortices provides evidence for a spontaneously broken U(1) symmetry and global interlayer phase coherence. This demonstration opens the door to phase-sensitive probes of exciton superfluidity, including interferometric measurements[57] and the exciton-condensate Josephson effect[58]. The ability to image and quantitatively characterize vortex matter further enables controlled studies of vortex dynamics, pinning, and melting in a purely electronic superfluid. Beyond the vortex lattice phase, graphene quantum Hall bilayers have been predicted to host more exotic correlated states, including fractional excitons[28] and supersolid phases[20,29]. Our real-space imaging approach provides a direct platform to search for such states, as well as for analogous excitonic phases in transition metal dichalcogenide heterostructures. Finally, spatially resolved tunnelling spectroscopy could probe defect-bound excitations of the condensate, potentially revealing the excitonic analogue of Yu–Shiba–Rusinov states[59].

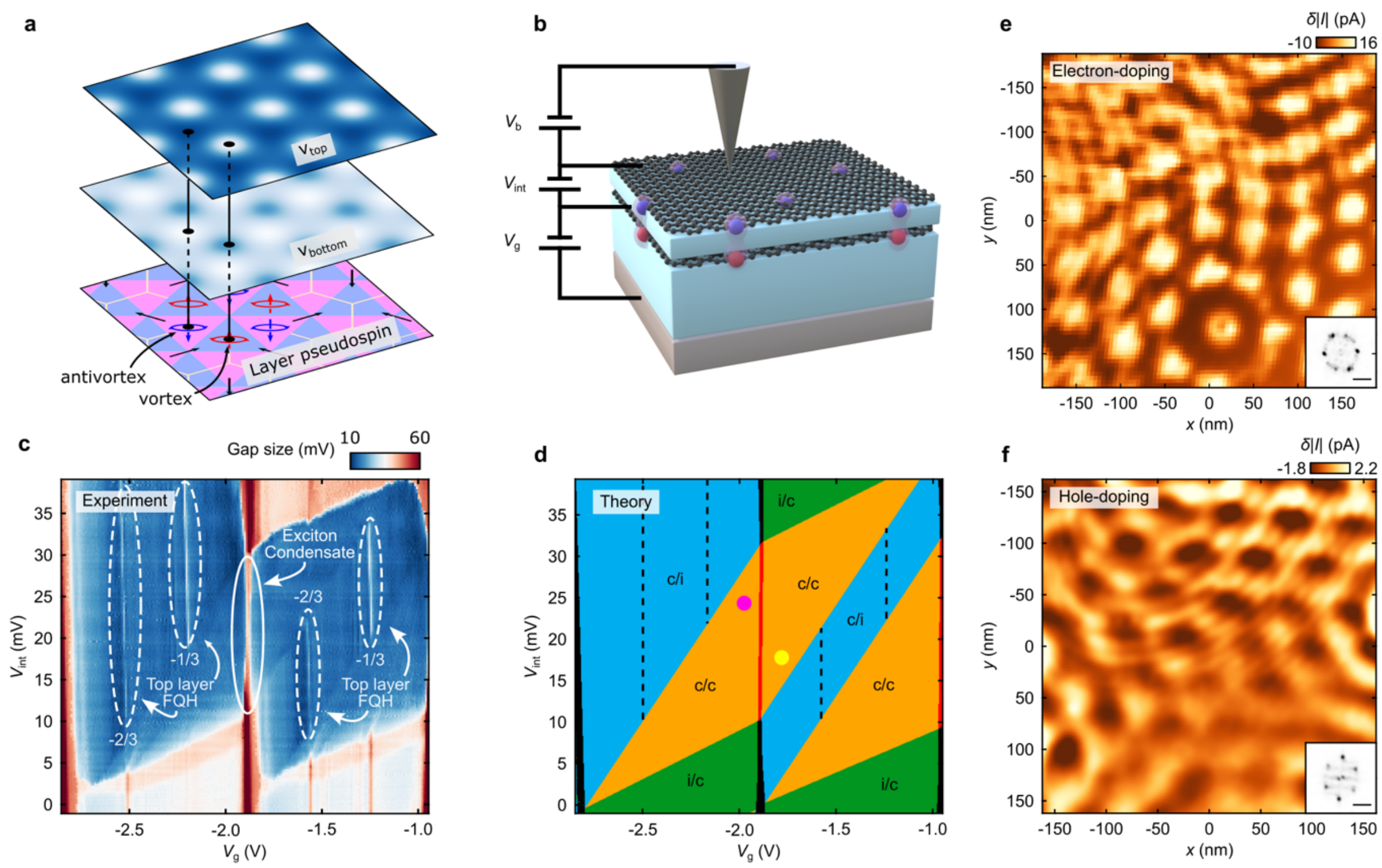

***Fig. 1: Exciton condensate and vortex lattice in double-layer graphene. a,*** *Illustration of a vortex lattice with the top two layers corresponding to the charge densities of the two graphene layers and the bottom layer corresponding to the local layer pseudospin, showing vortices (red) and antivortices (blue).* ***b,*** *Schematic of the device geometry with excitons where* $V_g$ *tunes the total filling factor in the two graphene layers,* $V_{int}$ *distributes charges between the two layers, and* $V_b$ *is the tunnelling bias.* ***c,*** *The size of the top layer's spectroscopic gap as a function of* $V_g$ *and* $V_{int}$*. See SI section 2.1 for a discussion of interlayer fractional quantum Hall states.* ***d,*** *A corresponding calculated electrostatic phase diagram with 4 combinations of the top and bottom layer being compressible and incompressible (labelled top/bottom). The exciton condensate is indicated by a vertical red stripe.* ***e-f,*** *Spatially resolved modulation in the tunnelling current magnitude* $\delta|I|$ *of a vortex lattice measured at the edge of the spectroscopic gap and T = 200 mK, B = 9 T. A Gaussian filter with* $\sigma$ *= 4 nm was applied to suppress effects from the graphene/hBN moiré. The insets show the moduli of structure factors of the image, with a scale bar indicating 0.02 nm*$^{-1}$*.*

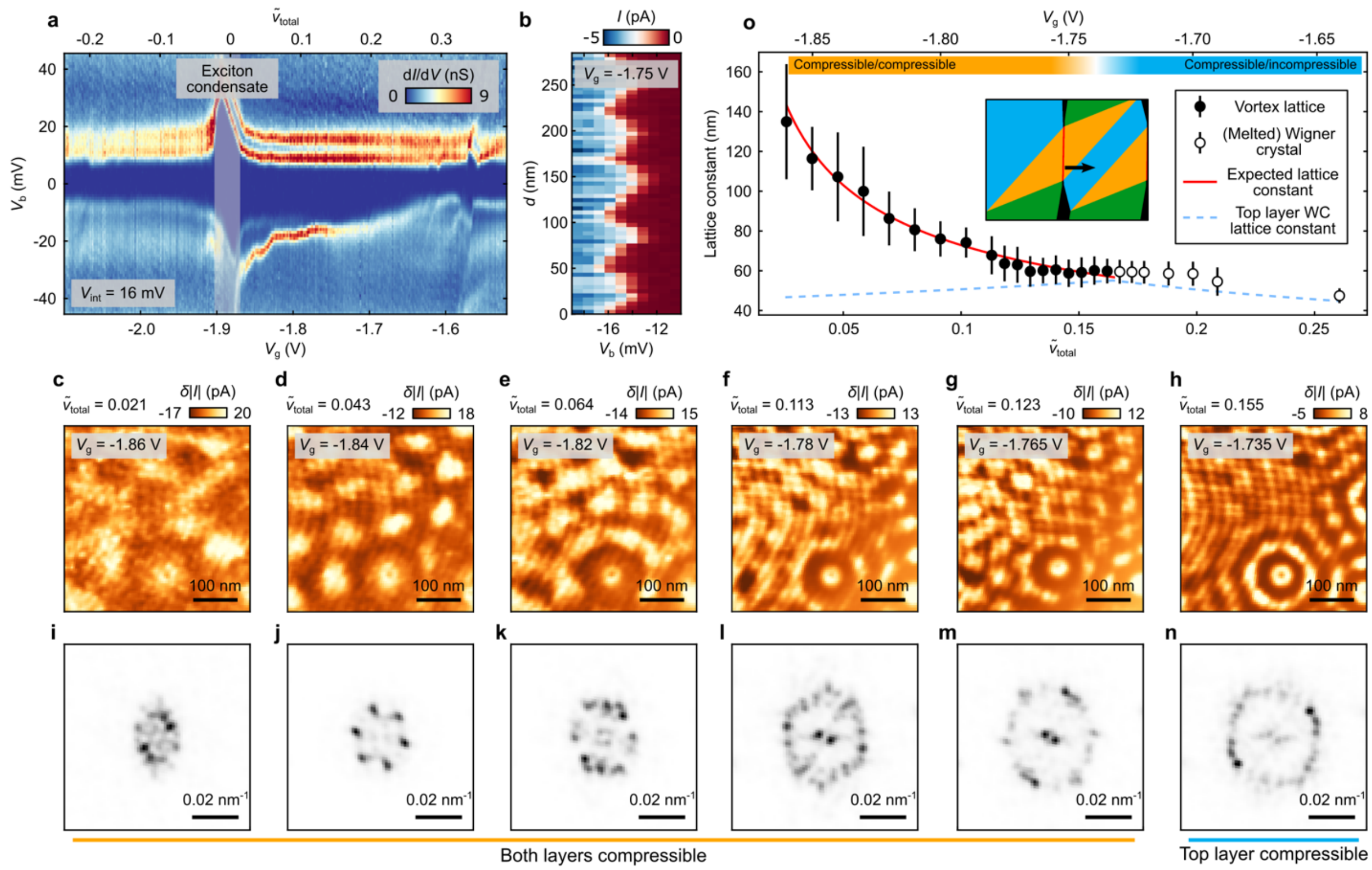

***Fig. 2: Evolution of an electron-doped vortex lattice as a function of backgate.*** ***a,*** *Tunnelling spectrum as a function of* $V_g$ *measured at* $V_{int}$ *= 16 mV.* ***b,*** *Spatial linecut of the measured tunnelling current* I *across several lattice sites.* ***c-h,*** *Spatially resolved modulation in the tunnelling current magnitude* $\delta|I|$ *(with* $\sigma$ *= 4 nm Gaussian filter) and the corresponding modulus of the structure factor* ***i-n*** *measured at various* $V_g$ *and fixed* $V_{int}$ *= 16 mV, with the total filling factor increasing from* $\nu_{total}$ *= -1.979 on the left to* $\nu_{total}$ *= -1.845 on the right.* ***o,*** *Lattice constant of the periodic pattern extracted from the structure factor (see Methods for details) as a function of* $V_g$*. The red curve represents a theoretically calculated lattice constant evolution of a vortex lattice and the blue curve of a top-layer Wigner crystal. At* $V_g \sim$ *-1.74 V, the bottom layer becomes incompressible, and all the charges reside in the top layer. An electrostatic phase diagram is shown in the inset, with the black arrow illustrating the backgate and interlayer voltages chosen for this measurement.*

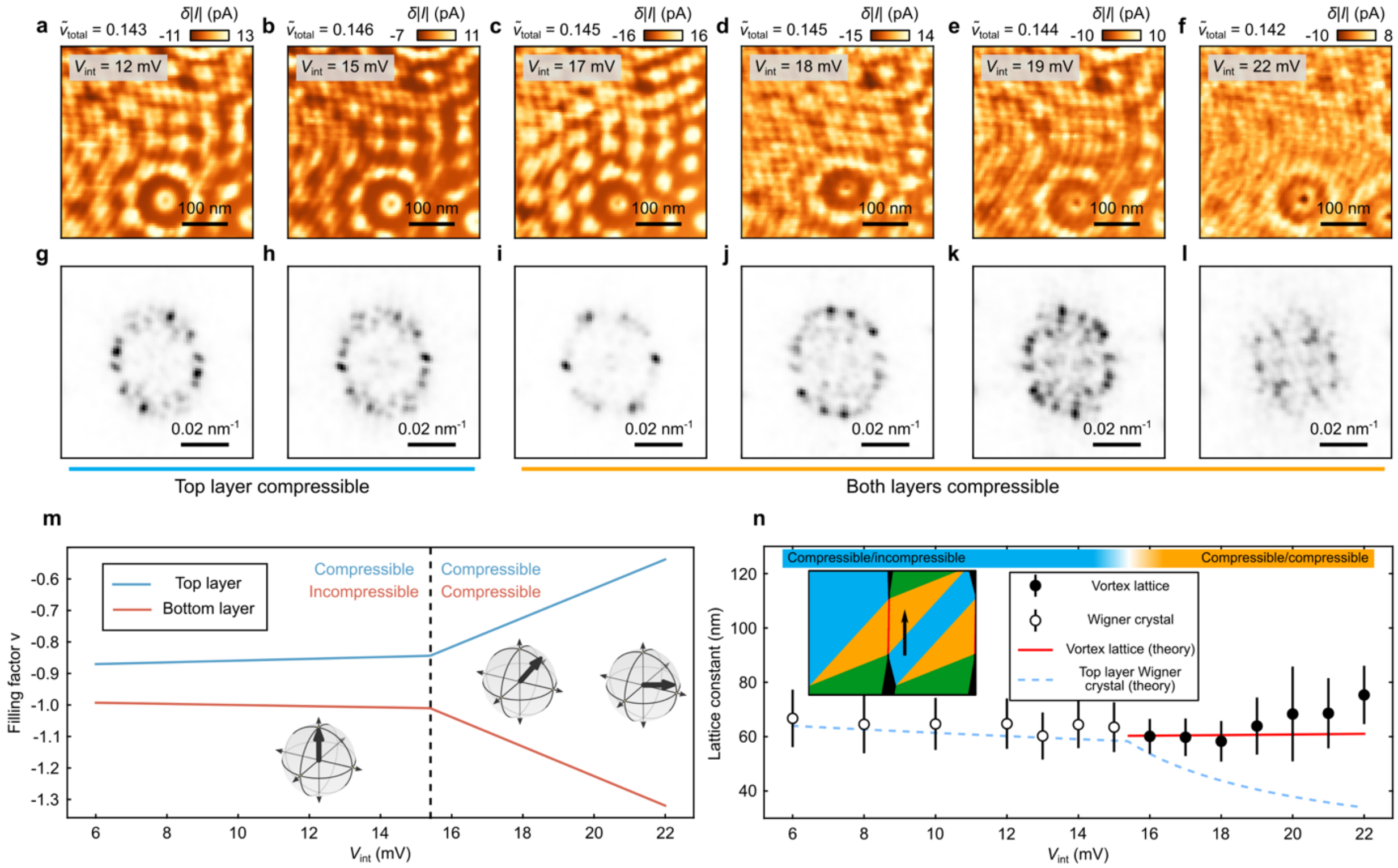

***Fig. 3: Evolution of a vortex lattice as a function of interlayer voltage. a-f,*** *Spatially resolved modulation in the tunnelling current magnitude δ|I|(with σ = 4 nm Gaussian filter) and their structure factors moduli* ***g-l*** *measured at various $V_{int}$ and fixed $V_g$ = -1.76 V, which corresponds to $\tilde{\nu}_{total}$ = 0.14.* ***m,*** *Evolution of the top and bottom layers' filling factors as a function of $V_{int}$. Layer pseudospin Bloch spheres illustrate the evolution of the layer polarization as $V_{int}$ is varied.* ***n,*** *Lattice constant of the periodic pattern extracted from the structure factor (see Methods for details) as a function of $V_{int}$ and fixed total filling factor. The red curve represents a theoretically calculated lattice constant evolution of a vortex lattice and the blue curve of a top-layer Wigner crystal. At $V_{int}$ < 16 mV, the bottom layer is incompressible and all the charges reside in the top layer. An electrostatic phase diagram is shown in the inset, with the black arrow illustrating the backgate and interlayer voltages chosen for this measurement.*

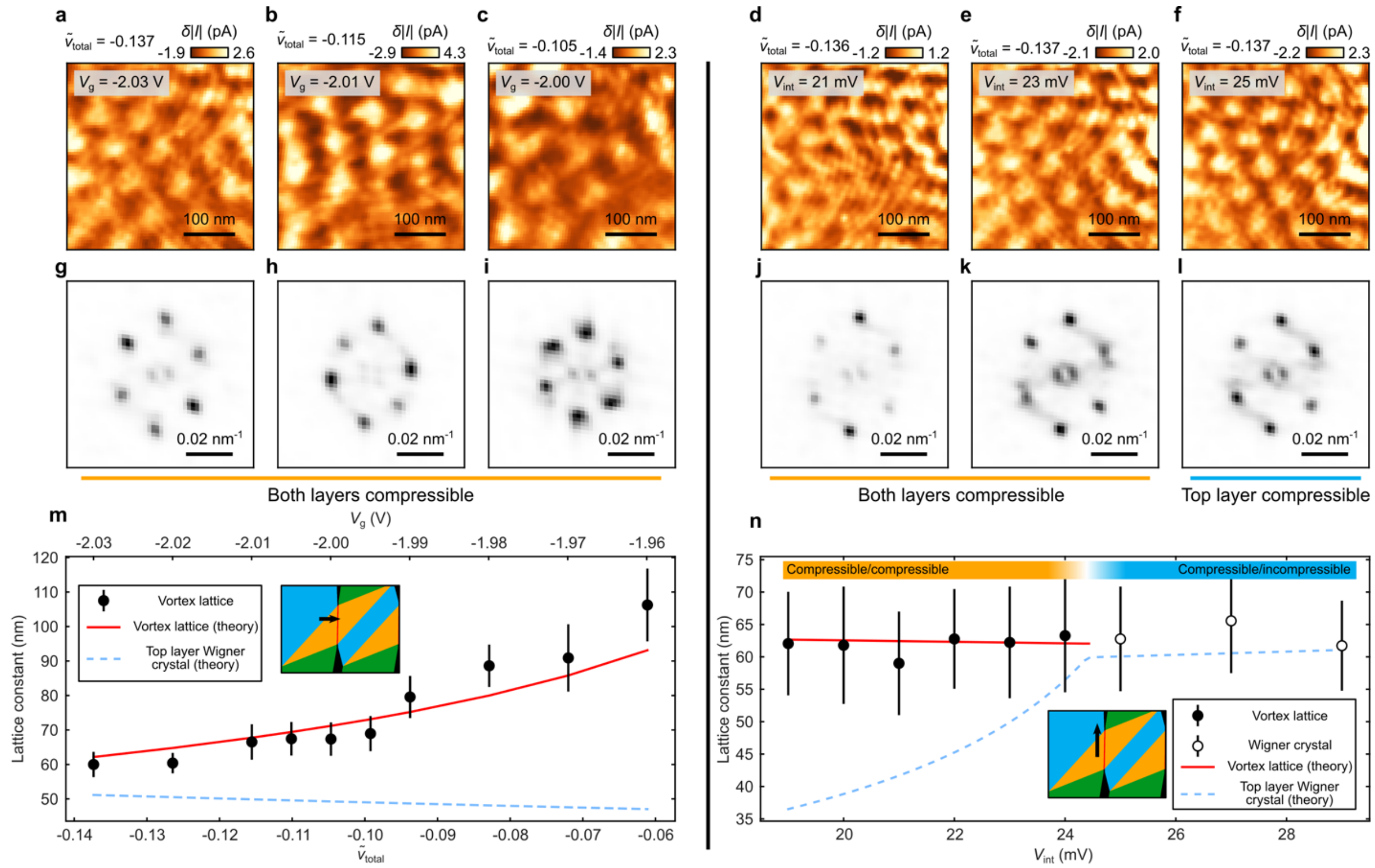

***Fig. 4: Evolution of a hole-doped vortex lattice as a function of backgate and interlayer voltage. a-c,*** *Spatially resolved modulation of the tunnelling current magnitude δ|I| (with σ = 4 nm Gaussian filter) and their structure factor moduli* ***g-i*** *measured at various* $V_g$ *and fixed* $V_{int}$ *= 23 mV.* ***m,*** *Lattice constant of the periodic pattern extracted from the structure factor (see Methods for details) as a function of* $V_g$ *at* $V_{int}$ *= 23 mV.* ***d-f,*** *Spatially resolved modulation of the tunnelling current magnitude δ|I| (with σ = 4 nm Gaussian filter) and their structure factor moduli* ***j-l*** *measured at various* $V_{int}$ *and fixed* $V_g$ *= -2.03 V, which corresponds to* $\tilde{\nu}_{total}$ *= -0.14.* ***n,*** *Lattice constant of the periodic pattern as a function of* $V_{int}$ *at* $V_g$ *= -2.03 V. In* ***m*** *and* ***n****, the red curve represents a theoretically calculated lattice constant evolution of a vortex lattice and the blue curve of a top-layer Wigner crystal.*

## Methods

### Device fabrication

The double-layer graphene devices were fabricated using a sliding disassembly dry transfer method[1]. The heterostructure was picked up with a large ~70-nm-thick hBN on top by a polyvinyl alcohol (PVA) handle, supported by a transparent tape on polydimethylsiloxane (PDMS) and then transferred onto a pre-patterned SiO2 (285 nm)/Si substrate with Au/Ti contacts (30 nm/3 nm). The PVA film was dissolved in HPLC water and the large hBN covering the sample was removed using the sliding disassembly method. Next, the devices were wirebonded onto the sample holder, transferred into the ultrahigh vacuum chamber and annealed at 600 °C overnight before being transferred into the STM.

### STM measurement

STM/STS measurements were performed using a homebuilt, ultrahigh-vacuum scanning tunnelling microscope[2] at phonon temperature $T_{ph}$ = 100 mK and effective temperature $T_{eff}$ = 200 mK and at a perpendicular magnetic field of 9 T. Tungsten scanning tunnelling microscope tips were prepared on a Cu(111) single crystal to match the work function of monolayer graphene and benchmarked using spectroscopy of the Shockley surface state of Cu(111).

The experiments were performed with the STM tip grounded and $V_b$ applied to the top graphene layer. Using a voltage adder, the voltage applied to the bottom graphene layer was $V_{int} + V_b$ and the backgate voltage was $V_g + V_{int} + V_b$. STS measurements were obtained through a standard lock-in detection method of the tunnelling current at 1541.2 Hz, induced by a sinusoidal modulation voltage added to $V_b$. Real-space $I(V_b)$ images were obtained using a grid spectroscopy measurement with the tip stabilizing at each point of the grid with a subsequent $I(V_b)$ acquisition. In this work, we acquired $I(V_b)$ maps instead of $dI/dV$ maps due to their higher signal-to-noise ratio and shorter integration time required. Tunnelling parameters for all measurements are provided in section 18 of the SI.

### Structure factor S(q) extraction

The structure factor of the I(r) map is defined as $S(q) = \langle I(q) * I(-q) \rangle$, where $I(q)$ is the Fourier transform of $I(r)$. The Fourier transform of the autocorrelation of $I(r)$ is equivalent to the modulus-squared fast Fourier transform map of the $I(r)$ map $|I(q)|^2$, which is the structure factor $S(q)$. In the main figures, we first obtain autocorrelation of $I(r)$ with zero padding, and perform the Fourier transform to obtain the structure factor $S(q)$.

**Determination of filling factor $\nu_i$ and the lattice constant $a_i$**

In the main figures, the top layer filling factor $\nu_{top}$, the bottom layer filling factor $\nu_{bot}$, and the total filling factor $\nu_{total}$ are determined from $V_g$ and $V_{int}$, using an electrostatic model which considers both geometric and quantum capacitances. The electrostatic model accurately reproduces the experimental phase diagram (details in SI section 2).

The lattice constant $a_i$ is obtained from the reciprocal relation ai · $q_j = \delta_{ij}$, where we have $1/|q| * (2/\sqrt{(3)})$ for the triangular / honeycomb lattice. $|q|$ is obtained from the structure factor $S(q)$. $S(q)$ is first converted into polar coordinates ($\rho$, $\theta$) and then averaged over $\theta$ (piezo calibration is performed at this step). $|q|$ is then extracted by performing a Gaussian fit (including a linear background to separate features of the lattice from the disorder potential typical at low $q$'s, see SI section 10 for more details) of the azimuthally averaged Bragg peaks. This determines the fitting error bar of the $|q|$. The 95% confidence interval of $|q|$ is then converted to the 95% confidence interval of the corresponding lattice constant $a_i$, from which we determine its error bar.

**Method References**

**Data Availability**

The data that support the findings of this study are available from the corresponding author upon request.

## Code Availability

The code that supports the findings of this study is available from the corresponding author upon request.

## Acknowledgements

We thank Y.-C. Tsui and J. Yu for helpful discussions.

## Funding Statement

The work at Princeton is primarily supported by the DOE-BES grant DE-FG02-07ER46419 and Gordon and Betty Moore Foundation's EPiQS initiative grant GBMF9469. Other support for work at Princeton is from the U.S. Army Research Office MURI project under grant number W911NF-21-2-0147 and W911NF261A052, ONR grant N000142412471, NSF-MRSEC through the Princeton Center for Complex Materials grant NSFDMR-2011750, NSF grant DMR-2312311, and NSF grant OMA-2326767. V. V. acknowledges fellowship support from the Princeton Center for Complex Materials. B. Z. and A. H. M. were supported by the NSF through the Center for Dynamics and Control of Materials, an NSF Materials Research Science and Engineering Center (MRSEC) under Cooperative Agreement DMR-2308817. K.W. and T.T. acknowledge support from the Elemental Strategy Initiative conducted by the MEXT, Japan, grant JPMXP0112101001, JSPS KAKENHI grants 19H05790 and JP20H00354.

## Author Contributions

R. L. L., J. L., V. V., and A. Y. designed the experiment. V. V., J. L., C.-L. C., and R. L. L. fabricated the devices used for the study. J. L., R. L. L., and V. V. carried out STM/STS measurements. R. L. L., J. L., and V. V. performed data analysis. B. Z. and A. H. M. performed the theoretical simulations. K. W. and T. T. synthesized the hBN crystals. All authors discussed the results and contributed to the writing of the manuscript.

**Competing Interests**

The authors declare no competing interests.

Correspondance and requests for materials should be addressed to Ali Yazdani yazdani@princeton.edu.